\documentclass[a4paper]{article}

\usepackage[dvipdfmx]{graphicx}

\usepackage{INTERSPEECH2020}
\usepackage{url}
\usepackage{hyperref}
\usepackage{here}

\newcommand{\sampleurl}{https://kentaro321.github.io/demo_DGP_MS_TTS/}

\usepackage{bm}

\newcommand{\muB}{\bm{\mu}}
\newcommand{\SigmaB}{\bm{\Sigma}}

\newcommand{\bB}{\mathbf{b}}

\newcommand{\hB}{\mathbf{h}}

\newcommand{\IB}{\mathbf{I}}

\newcommand{\mB}{\mathbf{m}}

\newcommand{\RB}{\mathbf{R}}
\newcommand{\rB}{\mathbf{r}}
\newcommand{\SB}{\mathbf{S}}

\newcommand{\UB}{\mathbf{U}}
\newcommand{\uB}{\mathbf{u}}

\newcommand{\WB}{\mathbf{W}}
\newcommand{\xB}{\mathbf{x}}
\newcommand{\XB}{\mathbf{X}}
\newcommand{\yB}{\mathbf{y}}
\newcommand{\YB}{\mathbf{Y}}

\newcommand{\ZB}{\mathbf{Z}}
\newcommand{\zeroB}{\mathbf{0}}

\newcommand{\LM}{\mathcal{L}}
\newcommand{\NM}{\mathcal{N}}
\newcommand{\SM}{\mathcal{S}}

\newcommand{\fzero}{{$f_\mathrm{o}~$}}
\newcommand{\fzeronosp}{{$f_\mathrm{o}$}}

\newcommand{\frm}{{\mathrm{f}}}

\title{Multi-speaker Text-to-speech Synthesis Using Deep Gaussian Processes}
\name{Kentaro Mitsui, Tomoki Koriyama, Hiroshi Saruwatari}
\address{
  The University of Tokyo, Japan
}
\email{[kentaro\_mitsui, tomoki\_koriyama, hiroshi\_saruwatari]@ipc.i.u-tokyo.ac.jp}

\begin{document}

\setlength{\abovedisplayskip}{2pt} 
\setlength{\belowdisplayskip}{2pt} 
\maketitle


\begin{abstract}
\vspace{-3pt}

Multi-speaker speech synthesis is a technique for modeling 
multiple speakers' voices
with a single model.
Although many approaches using deep neural networks (DNNs) have been proposed,
DNNs are prone to overfitting
when the amount of training data is limited.
We propose a framework for multi-speaker speech synthesis 
using deep Gaussian processes (DGPs);
a DGP is a deep architecture of Bayesian kernel regressions
and thus robust to overfitting.
In this framework, speaker information is fed to duration/acoustic models
using speaker codes.
We also examine the use of deep Gaussian process latent variable models (DGPLVMs).
In this approach, 
the representation of each speaker is learned simultaneously with other model parameters,
and therefore the similarity or dissimilarity of speakers 
is considered efficiently.
We experimentally evaluated two situations to investigate the effectiveness of 
the proposed methods.
In one situation, 
the amount of data from each speaker is balanced (speaker-balanced),
and in the other, the data from certain speakers are limited (speaker-imbalanced).
Subjective and objective evaluation results showed that
both the DGP and DGPLVM synthesize multi-speaker speech more effective
than a DNN in the speaker-balanced situation.
We also found that the DGPLVM outperforms the DGP significantly in the speaker-imbalanced situation.

\end{abstract}

\vspace{-7pt}
\section{Introduction}
\label{sec:intro}

With the development of machine learning in recent years,
text-to-speech (TTS) synthesis has a greater variety of applications 
than ever before.
Recent studies have shown that multi-speaker modeling,
a technique that models the voices of multiple speakers with a single model,
is effective for synthesizing multiple speakers' voices.
Multi-speaker modeling can benefit from 
multi-task learning~\cite{ruder2017overview},
which means this technique requires less training data
to achieve high-quality speech synthesis.

Statistical parametric speech synthesis (SPSS) is 
one possible method for multi-speaker speech synthesis.
Hidden Markov model (HMM)-based methods 
such as the average voice model~\cite{yamagishi2007average}
were widely used until the emergence of 
deep neural network (DNN)-based speech synthesis~\cite{zen13dnn}.
For multi-speaker modeling in DNN-based speech synthesis,
Fan et al. introduced a shared 
hidden-layer structure,
which shares the hidden-layer parameters of a DNN among different speakers,
and reported that this structure improved the quality of synthetic speech 
relative to the speaker-dependent DNNs~\cite{fan2015multi}.
Another successful method for multi-speaker modeling is based on speaker codes, 
which are the representation of speakers in a form
such as a one-hot vector or randomly assigned vector.
Luong et al. investigated the optimal form for 
speaker codes~\cite{luong2017adapting}.
The method proposed by Hojo et al. outperformed 
the shared hidden-layer structure
by feeding one-hot speaker codes to the hidden layers of a DNN~\cite{hojo2018dnn}.
In addition,
the method using speaker representation
has recently been applied to end-to-end speech synthesis frameworks,
and the method has achieved high speech quality~\cite{ping2017deep, jia2018transfer}.
However,
most of the DNN-based methods only consider data fitting while training,
and thus overfitting often becomes a problem.

In this paper, 
we focus on the SPSS framework using 
deep Gaussian processes (DGPs)~\cite{koriyama2019statistical}.
In this framework,
the relationship between linguistic features and phoneme durations or acoustic features
are modeled using DGPs~\cite{damianou2013deep}.
A DGP is a deep architecture of Bayesian kernel regressions,
so it can express complicated non-linear transformation with
a small number of hyperparameters.
Both data fitting and model complexity are considered
in the training of a DGP,
which makes the model less vulnerable to overfitting than a DNN.
Previous work has shown that
DGP-based TTS performs better than a feed-forward DNN 
for single-speaker modeling~\cite{koriyama2019statistical}.
However, the DGP's effectiveness for multi-speaker TTS is yet to be verified.

Therefore, 
we propose multi-speaker TTS based on DGP.
We introduce two methods:
one method using a general DGP and feeding one-hot speaker codes 
to its hidden layers,
similarly to the DNN-based method~\cite{hojo2018dnn};
and the other based on learning latent representation of speakers
using deep Gaussian process latent variable models (DGPLVMs)~\cite{damianou2013deep}.
The second method incorporates a GPLVM~\cite{titsias2010bayesian},
a Bayesian generative model
shown to be effective in prosody modeling~\cite{koriyama2019semi},
into the general DGP to obtain speaker representation.
The difference between DGPs and DGPLVMs is the representation of 
speaker similarity used for kernel regression.
A DGPLVM can explicitly express the similarity using the latent representation
whereas the speaker codes used in a general DGP cannot.
In addition,
the use of DGPLVM enables an analysis of speakers
in the latent space.

In the experimental evaluations, 
we investigate the performance of our methods in 
speaker-balanced and speaker-imbalanced situations.
In the speaker-imbalanced situation,
we first selected target speakers and 
used limited data for those speakers while training.
We conducted objective and subjective evaluations in both 
situations to evaluate the effectiveness of the proposed methods.
Experimental results showed that 
in the speaker-balanced situation,
both proposed methods improved the speech quality 
relative to the DNN-based method;
and in the speaker-imbalanced situation
where only five training utterances were used 
for the target speakers,
the DGPLVM improved naturalness and speaker similarity of synthetic speech.


\vspace{-7pt}
\section{Conventional methods}
\subsection{DNN-based multi-speaker TTS using speaker codes}
\label{sec:dnn}
\vspace{-4pt}

We give an overview of DNN-based multi-speaker TTS using speaker codes~\cite{hojo2018dnn}, 
a simple yet highly effective method within the 
SPSS framework.
Single-speaker models use only contextual factors as the inputs of 
duration/acoustic models,
but this method uses speaker codes as auxiliary inputs
to model speaker variation.
Here, speaker code $\SM$ is a one-hot vector representation of the speaker ID.
We apply linear transformation to this vector and add the result to hidden layers:
\begin{align}
    \hB^{\ell+1} = \varphi(\WB^{\ell+1}(\hB^\ell + \WB_\SM^\ell \SM) + \bB^{\ell+1})
\end{align}
where $\varphi(\cdot)$ is an activation function, 
$\hB^\ell$ is the component of the $\ell$-th hidden layer, 
$\WB^\ell$ and $\WB_\SM^\ell$ are the connection weight of the hidden layers and speaker codes, respectively,
and $\bB^\ell$ is the bias.
Training is conducted by minimizing the mean squared error 
between the natural and generated acoustic features.


\vspace{-5pt}
\subsection{DGP-based speech synthesis}
\label{sec:dgp}

In the DGP-based speech synthesis framework~\cite{koriyama2019statistical},
a DGP model takes linguistic features as inputs
and predicts phoneme durations or acoustic features.
A DGP is a model defined as a cascade of 
Gaussian process regressions (GPRs).

GPRs model the relation between 
input $\xB$ and output $y$ as:
\begin{align}
    y &= f(\xB) + \epsilon \\
    f &\sim \mathcal{GP}(m(\xB), k(\xB, \xB'))
\end{align}
and infer the posterior distribution $p(y_*|\xB_*, \XB, \yB)$ against
the new input $\xB_*$ by using the training data $(\XB, \yB)$.
Here $\epsilon$ is random noise, and 
$m(\xB)$ and $k(\xB, \xB')$ are mean and kernel functions, respectively.
We consider multiple GPRs when the output is multidimensional.

Although a single GPR can represent complicated non-linear functions,
its expressiveness is limited by the kernel function.
A DGP overcomes this limitation by stacking multiple GPRs;
this method is based on the assumption that 
the overall function $f$ can be decomposed into multiple functions 
in the following manner:
\begin{align}
    f = f^{L+1} \circ f^{L} \circ \dots \circ f^{1}
\end{align}
where $L$ is the number of hidden layers,
and each function $f^\ell$ is a sample of a Gaussian process.
An approximation technique called
doubly stochastic variational inference~\cite{salimbeni2017doubly}
is used in this framework, 
so training is conducted by maximizing the evidence lower bound (ELBO) of log marginal likelihood:
\begin{align}
    \log p(\YB) &\geq 
    \frac{1}{N_s} \sum_{j=1}^{N_s} \sum_{i=1}^N 
    \Biggl\{ \sum_{d=1}^{D_{L+1}} \mathbb{E}_{q\left(\frm_{i,j}^{d}\right)}
    \left[ \log p\left(y_i^d | \frm_{i,j}^{d}\right) \right]
    \notag \\ &~~~~~~
    - \frac{N_s}{N} \sum_{\ell=1}^{L+1} \mathrm{KL}\left[ q(\UB^{\ell}) \| p(\UB^{\ell} | \ZB^\ell) \right] \Biggr\} \triangleq \LM_1
    \label{eq:dsvi_elbo}
\end{align}
where $N$, $N_s$ are the number of training data and Monte Carlo samples,
respectively,
and $D_\ell$ is the dimensionality of the output of the $\ell$-th GPR.
$y_i^d$ is the $d$-th dimension of the $i$-th observed output $\yB_i$,
and $\frm_{i, j}^d$ represents the corresponding latent function
predicted from the $j$-th sample point.
$\ZB^\ell$ and $\UB^\ell$ denote the inducing inputs and outputs,
respectively,
which are sparse representations of input and output data.
While $\ZB^\ell$ is a model parameter by itself,
$\UB^\ell$ itself is not a parameter but a random variable,
in which we impose 
$q(\UB^\ell) = \Pi_{d=1}^{D_\ell}q(\uB^{\ell, d}) 
= \Pi_{d=1}^{D_\ell}\NM(\uB^{\ell, d}; \mB^{\ell, d}, \SB^{\ell, d})$
and regard mean $\mB^{\ell, d}$ and variance $\SB^{\ell, d}$ as model parameters
for each layer $\ell$ and dimension $d$.


\vspace{-5pt}
\section{DGP-based multi-speaker TTS using speaker codes}

\begin{figure}[t]
\centering
  \includegraphics[width=0.5\hsize]{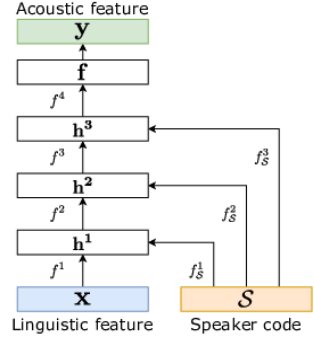}
  \vspace{-5pt}
  \caption{Architecture of DGP-based acoustic model for multi-speaker TTS with 
  three hidden layers.}
  \label{fig:dgp}
\vspace{-12pt}
\end{figure}

We introduce the model architecture shown in Fig.~\ref{fig:dgp}
to apply the DGP-based speech synthesis framework~\cite{koriyama2019statistical}
to multi-speaker TTS.
In this architecture,
speaker IDs are represented using one-hot speaker codes
in a manner similar to the DNN-based method
described in Section~\ref{sec:dnn}.
We apply a single-layer GPR to these speaker codes 
before feeding them to the hidden layers.
Therefore,
the values of the $\ell$-th hidden layer $\hB^\ell$ can be written as:
\begin{align}
    \hB^\ell = f^\ell(\hB^{\ell-1}) + f_\SM^\ell(\SM)
\end{align}
where $\SM$ denotes the speaker code,
$f^\ell$ is the $\ell$-th GPR in the DGP (hereinafter called the hidden GP),
and $f_\SM^\ell$ is the $\ell$-th GPR to transform speaker codes 
(hereinafter called the speaker GP).
Speaker GPs have inducing inputs $\ZB_\SM^\ell$ and corresponding outputs
$\UB_\SM^\ell$ as well as hidden GPs,
so we must optimize these parameters jointly with other model parameters.
This can be done by maximizing the new ELBO:
\begin{align}
    \LM_2 &= \LM_1 - \sum_{\ell=1}^{L} \mathrm{KL}\left[
    q(\UB_\SM^{\ell}) \| p(\UB_\SM^{\ell} | \ZB_\SM^\ell) \right].
    \label{eq:msdgp_elbo}
\end{align}

\vspace{-5pt}
\section{DGPLVM for multi-speaker TTS}
\label{sec:dgplvm}

In this section,
we propose another approach for multi-speaker TTS
using a DGPLVM~\cite{damianou2013deep}.
The DGP-based approach illustrated in the previous section is straightforward,
but because one-hot speaker codes are orthogonal to each other between speakers,
we cannot fully make use of the similarity or dissimilarity of speakers.
In the DGPLVM-based approach, 
we aim to 
utilize speaker similarity for multi-speaker TTS.

We express $K$ speakers by using latent variable 
$\RB = (\rB_1, ..., \rB_K)$, 
and use the latent variable as the input of function 
$f^\ell$ as follows: 
\begin{align}
    f^\ell \sim \mathcal{GP}(m(\xB, \rB_k), k([\xB^\top, \rB_k^\top]^\top , [\xB'^\top, \rB_{k'}^\top]^\top)) .
\end{align}
From Bayes' theorem, 
the distribution of $\rB_k$ conditioned on 
input $\xB$ and output $\yB$ can be written as:
\begin{align}
    p(\rB_k|\xB, \yB) 
    &\propto p(\yB|\xB, \rB_k)p(\rB_k).
    \label{eq:r_posterior}
\end{align}
When we consider acoustic modeling, the left-hand side of (\ref{eq:r_posterior}) is conditioned not only on linguistic feature $\xB$ but also on acoustic feature $\yB$.
Since the kernel function uses latent variable $\rB_k$ as input, $\rB_k$ is learned to express the similarity of acoustic features among different speakers.
We assign a prior given by the standard normal distribution to $\rB_k$:
\begin{align}
    p(\rB_k) = \NM(\rB_k; \zeroB, \IB).
\end{align}
Also, 
we consider the latent variable for $k$-th speaker $\rB_k$ 
to have a variational distribution
\begin{align}
    q(\rB_k) = \NM(\rB_k; \muB_k, \SigmaB_k)
\end{align}
where $\muB_k$ is a mean vector and
$\SigmaB_k$ is a diagonal covariance matrix.
This latent variable is fed to an arbitrary hidden layer of the DGP.
In this case the ELBO of $\log\int p(\YB|\RB)p(\RB)d\RB$ is written as:
\begin{align}
    \LM_3 = \LM_1 - \sum_{k=1}^K \mathrm{KL}\left[q(\rB_k) \| p(\rB_k) \right] .
\end{align}

\begin{figure}[t]
\centering
  \includegraphics[width=0.5\hsize]{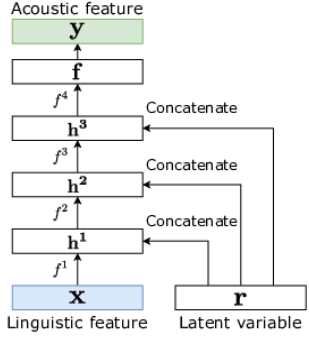}
  \vspace{-5pt}
  \caption{Architecture of DGPLVM-based acoustic model for multi-speaker TTS with three hidden layers.}
  \label{fig:dgplvm}
\vspace{-10pt}
\end{figure}


\vspace{-5pt}
\section{Experiments}
\subsection{Experimental conditions}

We used JVS corpus~\cite{takamichi2019jvs},
which is comprised of speech data from 100 Japanese speakers,
49 males and 51 females.
Speech waveforms were downsampled to 16~kHz.
This corpus contained 100 parallel utterances (parallel100) 
and 30 non-parallel utterances (nonpara30) from each speaker.
For the speaker-balanced situation,
the training set consisted of 
all the non-parallel and 85 of the 100 parallel utterances
from each speaker,
and the test set consisted of the remaining 15 parallel utterances 
from each speaker.
For the speaker-imbalanced situation,
four speakers, 
two males and two females,
were selected as target speakers;
for these speakers,
only five non-parallel utterances were used in training.
To avoid low speech quality for the target speakers,
we used an oversampling technique~\cite{luong2019training}
and sampled each utterance of each target speaker 20 times.
The target speakers were selected on the basis of 
subjective speaker similarity~\cite{Saito2019}.
Specifically, 
we defined the speaker who had the largest median of similarity score between other speakers, in other words who had many similar speakers,
as {\it male/female similar} ({\tt MS/FS}),
and the opposite ones as {\it male/female dissimilar} ({\tt MD/FD}).
The test set consisted of 15 parallel utterances from the four target speakers.

The input linguistic features of the duration model were 531-dimensional vectors
containing contextual factors such as phoneme, accent, and part of speech,
which were automatically estimated from texts using Open JTalk~\cite{oura2010japanese}.
We added a four-dimensional frame index to these linguistic features and
used them as the input of the acoustic model.
The output of the duration model was a one-dimensional phoneme duration.
The acoustic features, i.e. the output of the acoustic model, were 187-dimensional vectors
comprised of 0--59th mel-cepstrum,
log \fzeronosp,
coded aperiodicity and their $\Delta$, $\Delta^2$,
followed by voiced/unvoiced flags.
These acoustic features were extracted every 5~ms
using WORLD~\cite{morise2016world} (D4C edition~\cite{morise2016d4c}).
We normalized input features to range $[0.01, 0.99]$
and output features to zero-mean and unit variance.

The DGP duration model had 2 hidden layers,
with the dimensionality of each layer set to 32.
The acoustic model had 5 hidden layers,
and the dimensionality of each layer was 128.
The number of inducing points was set to 1024 for 
hidden GPs and 8 for speaker GPs.
We used ArcCos kernel~\cite{cho2009kernel} as a kernel function of GPs.
The inducing inputs of each GP were initialized randomly with
the standard normal distribution.
The variational distributions of inducing outputs $q(\uB^{\ell, d})$ of all GPs 
except the last hidden GP $f^{L+1}$ were initialized with a
Gaussian distribution with zero mean and variance $10^{-6}$,
while that of $f^{L+1}$ had unit variance.

The DGPLVM had similar settings to the DGP model.
However,
it does not have speaker GPs
and thus the total number of model parameters was reduced.
The variational distributions of latent variables $q(\rB_k)$ were
initialized randomly with Gaussian distribution with zero mean and variance $10^{-4}$.

\begin{figure}[t]
\centering
  \includegraphics[width=\hsize]{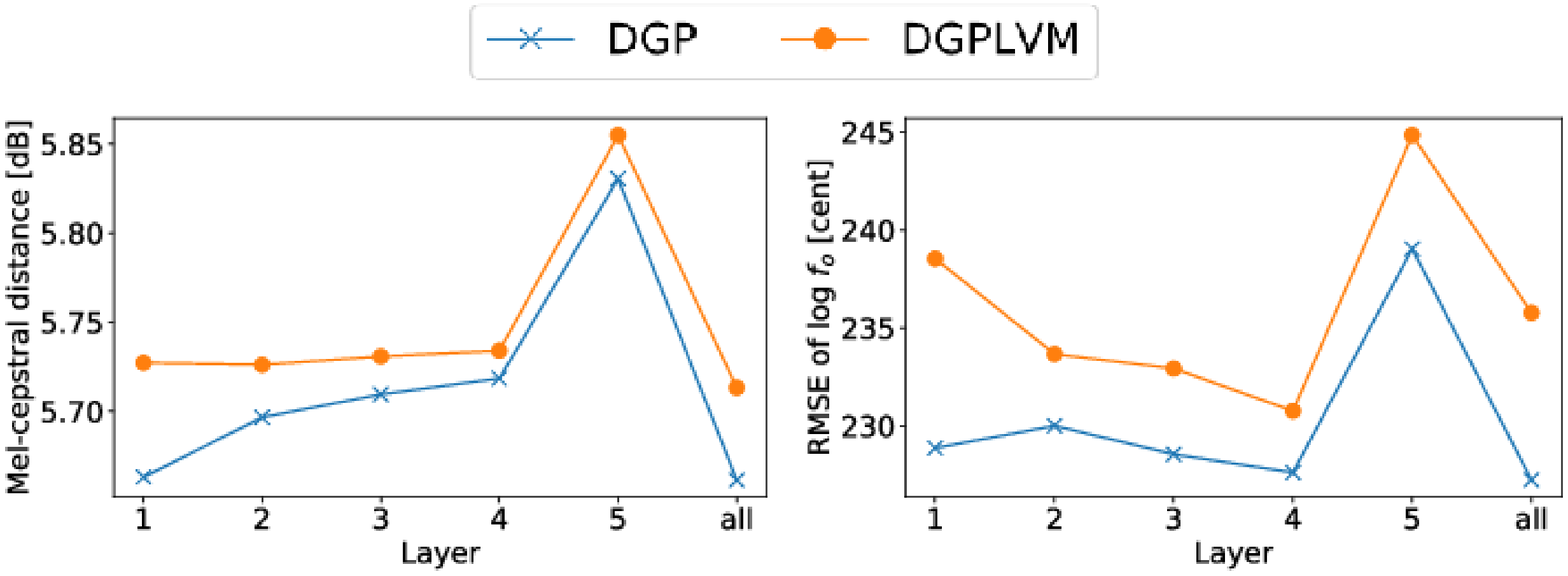}
  \vspace{-10pt}
  \caption{Objective evaluation results for DGP and DGPLVM with different layers to feed speaker information.}
  \label{fig:layer}
\vspace{-10pt}
\end{figure}

We trained the models by mini-batch optimization with the batch size set to 1024,
using Adam~\cite{kingma14adam}
whose learning rate was 0.01.
For the conventional DNN model, 
we followed the previous work~\cite{hojo2018dnn} and 
set the numbers of hidden layers to 2 and 5 for
duration and acoustic models, respectively,
the number of hidden units to 1024,
and the learning rate of Adam to $10^{-4}$.
Training was conducted up to 50 epochs for the DGP/DGPLVM and 
100 epochs for the DNN.


\vspace{-5pt}
\subsection{Objective evaluation}

We compared the quality of synthetic speech in terms of 
distortions between the original and synthetic speech parameters.
As evaluation metrics,
we used the root mean squared error (RMSE) of phoneme durations (DUR)
for duration models,
and mel-cepstral distance (MCD) and
RMSE of log \fzero (F0)
for acoustic models.

We first focused on the speaker-balanced situation 
and investigated the effect of 
model architecture on the performance of acoustic modeling.
For the DGP, 
we fed the speaker code $\SM$ to a certain layer 
(the first, second, third, fourth, or fifth layer) 
or all hidden layers of the acoustic model.
In the same way, for the DGPLVM, 
we fed the latent speaker variable $\rB_k$ to different layers.
Here the dimensionality of $\rB_k$ was set to three.
The results are shown in Fig.~\ref{fig:layer}.
Although feeding speaker information only to the last hidden layer 
increased the acoustic distortion, 
the differences among other settings were relatively small.
In the following experiments, 
we adopted the {\it all} settings for both the DGP and DGPLVM.

Next, we investigated the performance of the DGPLVM 
with different dimensionality of $\rB_k$.
We set the dimensionality of $\rB_k$ to 2, 3, 16, and 64. 
The results are shown in Table ~\ref{table:dimensionality}.
While higher dimensionality led to smaller distortions
in the speaker-balanced situation,
the results in the speaker-imbalanced situation were the opposite; 
lower dimensionality led to better results, 
and a dimensionality of three was optimal.
This is possibly because latent speaker space becomes dense
with low-dimensional speaker representation,
and voice models of similar speakers are efficiently accounted for
when synthesizing the target speaker's voice.
We set the dimensionality of $\rB_k$ to 64 for the speaker-balanced situation 
and 3 for the speaker-imbalanced situation in the following experiments.

Finally, we compared the performance of the conventional DNN, 
proposed DGP, and DGPLVM.
In the speaker-balanced situation, 
all models yielded similar MCD, 
while the proposed DGP/DGPLVM showed better F0 and DUR than the DNN.
In the speaker-imbalanced situation,
DNN was the best in terms of MCD
and DGPLVM was the best in terms of F0 and DUR.

\begin{table}[t]
\centering
\scriptsize
\renewcommand{\arraystretch}{1.2}
\caption{Objective evaluation results for DGPLVM with different dimensionality of latent speaker variable $\rB_k$. MCD: mel-cepstral distance [dB], F0: RMSE of log \fzero [cent].}
\label{table:dimensionality}
\vspace{-5pt}
\begin{tabular}{c|cc|cc} \hline 
& \multicolumn{2}{c|}{Speaker-balanced} & \multicolumn{2}{c}{Speaker-imbalanced} \\ \hline
Dimensionality & MCD & F0 & MCD & F0\\ \hline \hline
2 & 5.72 & 235 & 6.24 & 280 \\
3 & 5.71 & 236 & \bf{6.15} & \bf{264} \\
16 & \bf{5.65} & 233 & 6.28 & 285 \\ 
64 & \bf{5.65} & \bf{228} & 6.31 & 282 \\ \hline
\end{tabular}
\vspace{-5pt}
\end{table}

\begin{table}[t]
\centering
\scriptsize
\renewcommand{\arraystretch}{1.2}
\caption{Comparison of DNN, DGP and DGPLVM in terms of
MCD: mel-cepstral distance [dB], F0: RMSE of log \fzero [cent], 
and DUR: RMSE of phoneme duration [ms].}
\label{table:obj_eval}
\vspace{-5pt}
\begin{tabular}{c|ccc|ccc} \hline 
& \multicolumn{3}{c|}{Speaker-balanced} & \multicolumn{3}{c}{Speaker-imbalanced} \\ \hline 
Method  & MCD & F0 & DUR & MCD & F0 & DUR \\ \hline \hline
DNN & 5.66 & 239 & 25.6 & \bf{5.96} & 271 & 28.0 \\
DGP & 5.66 & \bf{227} & 25.4 & 6.29 & 280 & 27.7 \\ 
DGPLVM & \bf{5.65} & 228 & \bf{24.9} & 6.15 & \bf{264} & \bf{27.6} \\ \hline
\end{tabular}
\vspace{-15pt}
\end{table}


\vspace{-5pt}
\subsection{Subjective evaluation}

We conducted listening tests to subjectively evaluate the speech quality
in terms of naturalness and speaker similarity\footnote{Synthetic speech samples are available at \href{\sampleurl}{\url{\sampleurl}}.}.
The naturalness of synthetic speech was evaluated by preference A/B test,
and speaker similarity was evaluated by XAB test.
We compared two pairs: DNN--DGP and DGP--DGPLVM
in the speaker-balanced/imbalanced situations.
Thirty crowdsourced listeners participated in each of the evaluations, 
and each listener evaluated ten speech samples. 
The original speech of the target speaker was used 
as the reference X in the XAB tests.

\begin{figure}[t]
\centering
\vspace{-5pt}
  \includegraphics[width=0.9\hsize]{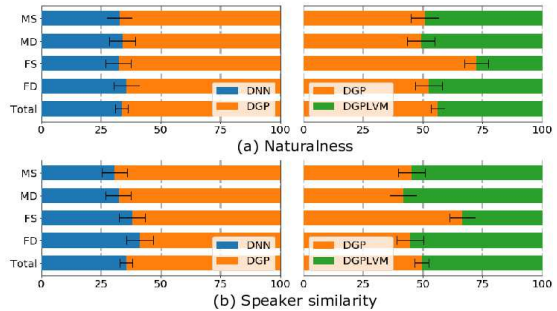}
  \vspace{-8pt}
  \caption{Subjective evaluation results with 95\% confidence intervals in speaker-balanced situation.}
  \label{fig:subjective_eval_balanced}
\vspace{-5pt}
\end{figure}

\begin{figure}[t]
\centering
\vspace{-5pt}
  \includegraphics[width=0.9\hsize]{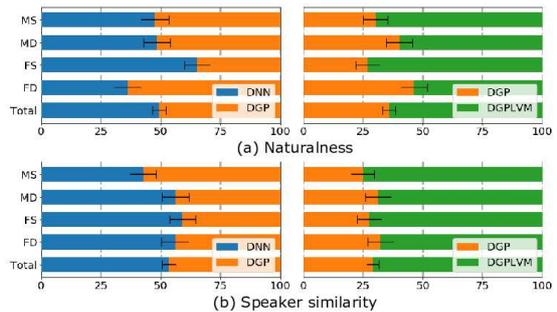}
  \vspace{-8pt}
  \caption{Subjective evaluation results with 95\% confidence intervals in speaker-imbalanced situation.}
  \label{fig:subjective_eval_imbalanced}
\vspace{-10pt}
\end{figure}

The results are shown in Figs.~\ref{fig:subjective_eval_balanced} and 
\ref{fig:subjective_eval_imbalanced}.
In the speaker-balanced situation,
the scores of both naturalness and speaker similarity were higher for all speakers for the DGP than for the DNN.
Although both scores of {\tt FS} were lower in the DGPLVM than in the DGP 
due to duration errors,
the scores of the remaining three speakers were comparable in DGP--DGPLVM.
Collating these results with those of the objective evaluation,~\fzero 
seems to have the greatest effect on 
naturalness and speaker similarity.

In the speaker-imbalanced situation,
there was no significant difference between the DNN and DGP in total,
though we observed larger acoustic feature distortions for the DGP 
in the objective evaluation.
The naturalness of the DGPLVM for {\tt MS} and {\tt FS} were 
significantly higher than those of the DGP.
In addition, the speaker similarity of those speakers were 
slightly higher than those of the other speakers in the DGPLVM.
From these results, 
we infer that the DGPLVM can beneficially utilize similar speakers
using the learned latent speaker representation.

\vspace{-5pt}
\subsection{Latent speaker representation learned by DGPLVM}

\begin{figure}[H]
\vspace{-5pt}
\centering
  \includegraphics[width=\hsize]{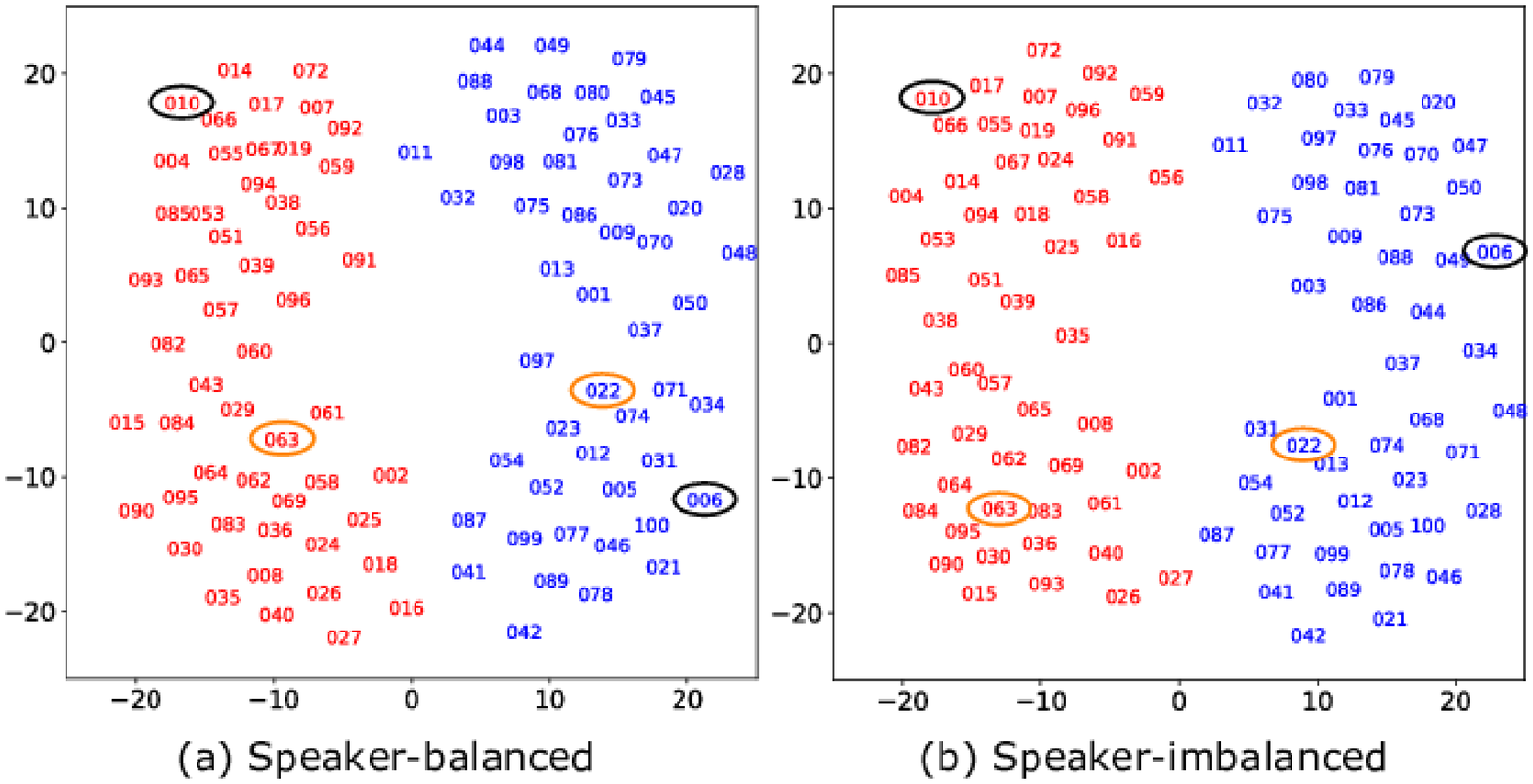}
  \vspace{-10pt}
  \caption{Latent speaker representation learned by DGPLVM in (a) speaker-balanced situation and (b) speaker-imbalanced situation. Red and blue numbers indicate female and male speakers, respectively. Orange and black circles indicate the {\rm similar} and {\rm dissimilar} speakers, respectively.}
  \label{fig:latent}
\vspace{-12pt}
\end{figure}

The latent speaker representation after training the DGPLVM is shown in Fig.~\ref{fig:latent}.
Here, the dimensionality of $\rB_k$ is set to two
for ease of visualization.
We found that 
male and female speakers were clearly separated,
{\it similar} speakers ({\tt MS}: 022 and {\tt FS}: 063) were embedded inside of the cluster 
while {\it dissimilar} speakers ({\tt MD}: 006 and {\tt FD}: 010) were embedded outside,
and
speakers embedded closely in the speaker-balanced situation were also closely embedded in the speaker-imbalanced situation.
These results indicate that the learned latent speaker representation
expresses the similarity or dissimilarity of speakers as expected.

\vspace{-5pt}
\section{Conclusions}

We have proposed multi-speaker TTS based on the DGP.
We found that with one-hot speaker codes,
the use of the DGP can improve naturalness and 
speaker similarity of multi-speaker speech
relative to the DNN.
We also introduced the DGPLVM-based multi-speaker TTS framework,
in which speaker representation is treated as a latent variable
and jointly learned with other model parameters.
The experimental results showed that 
the DGPLVM-based approach is especially effective
when the amount of training data from a certain speaker is 
highly limited.
For future work, 
we will compare our DGPLVM-based method with other latent-space-based methods such as variational autoencoder~\cite{kingma2013auto}.
We also plan to compare the performance of the proposed methods 
with recent end-to-end approaches.


\vspace{-5pt}
\section{Acknowledgements}

This work was supported by JSPS KAKENHI Grant Number JP19K20292.

\clearpage

\bibliographystyle{IEEEtran}

\bibliography{refs}

\end{document}